\documentstyle[11pt,paspconf,epsf,psfig]{article}
\def\kms{km s$^{-1}$}
\def\h{$h^{-1}$}
\begin{document}
\title{The Convergence Depth of the Local Peculiar Velocity Field}
\author{Daniel A. Dale}
\affil{Infrared Processing and Analysis Center, California Institute of Technology, Pasadena, CA 91024, USA}
\author{Riccardo Giovanelli}
\affil{Department of Astronomy, Cornell University, Ithaca, NY 14853, USA}

\begin{abstract}
We have obtained Tully-Fisher (TF) measurements for some 3000 late-type 
galaxies in the field and 
in 76 clusters 
distributed throughout the sky between $\sim$10 and 200\h\ Mpc. The cluster data are applied to the construction of an $I$ band TF template, resulting in a relation with a scatter of $\sim$0.35 magnitudes and a zero-point accurate to 0.02 magnitudes.  Peculiar motions are computed by referral to the template relation, and the distribution of line-of-sight cluster peculiar motions is presented.

The dipole of the reflex motion of the Local Group of galaxies 
with respect 
to galaxies with measured peculiar velocity
converges to the CMB dipole within less than 6000 \kms. 
The progression of this convergence is well illustrated when the reflex motion is referred to a well--distributed sample of field galaxies, and it is maintained when the reflex motion is
referred to the reference frame constituted by the distant clusters in our sample.  The field and cluster samples exhibit bulk motion amplitudes of order 200 \kms\ or smaller.    

Finally, we apply our sample of
 cluster peculiar velocities to a test of the putative Hubble bubble recently claimed by Zehavi and coworkers.  In contrast to their findings, our data support a relatively quiescent Hubble flow beyond $\sim$35\h\ Mpc.
\end{abstract}

\section{The SFI, SCI and SCII Surveys}
The results discussed in this report derive from peculiar velocity measurements obtained for three samples of spiral galaxies, based on the Tully-Fisher (TF) technique which combines $I$ band photometry and either 21 cm or H$\alpha$ long-slit spectroscopy.  The three samples are: (a) SFI, which includes 1631 field galaxies out to $cz\simeq 
6500
$ \kms\ (Haynes et al. 
1998a,b); 
(b) SCI, which includes 782 galaxies in the fields of 24 clusters within $cz\sim 9000$ \kms\ (Giovanelli et al. 1997a,b); and (c) SCII, which includes 522 galaxies in the fields of 52 clusters between $cz\simeq 5,000$ and $20,000$ \kms\ (Dale et al. 1997, 1998, 1999b,c).  For 
the merged cluster sample
SCI+SCII, there is an overlap of: 49 clusters with the Lauer \& Postman (1994) cluster sample; 27 clusters with the SMAC sample (Hudson et al. 1999); and 22 clusters with the EFAR sample (Wegner et al. 1996).

Detailed criteria for the construction of the samples are given elsewhere:
for SCI see Giovanelli et al. (1997a); sample characteristics of SCI are
given in Giovanelli et al. (1994); for SCII see Dale et al. (1999c).
Driving concerns for the selection of each sample were: dense sampling
of the local velocity field (SFI), high quality determination of the
TF template relation (SCI) and reliable recovery of the dipole reflex
motion of the Local Group (SCI+SCII).  The latter sample was constructed
with the aid of Monte Carlo realizations of the cluster distribution
and velocity field, utilizing mock samples from $N$-body simulations obtained
under a variety of cosmological scenarios, with the collaboration of
Stefano Borgani.

\section{The $I$ Band Tully-Fisher Template Relation}
Tully-Fisher distances, 
${\rm c}z_{\rm tf}$,
and therefore peculiar velocities
\begin{equation}
V_{\rm pec} = {\rm c}z - {\rm c}z_{\rm tf},
\end{equation}
are computed by 
reference
with a fiducial {\it template} relation which must be observationally derived.  The template relation defines the rest reference frame, against which peculiar velocities are to be measured.  The importance of accurately calibrating such a tool cannot be overemphasized.  It is possible that the discrepancies between different claims of bulk motions may be partly related to insufficiently well determined template relations.

For an assumed linear TF relation we need to determine two main parameters: a slope and a magnitude zero point.  The slope of the TF relation is best determined by a sample that maximizes the dynamic range in absolute magnitude $M_I$ and disk rotational velocity width $W$, i.e. one that preferentially includes nearby objects.  The proximity of the SCI sample provides the 
dynamic range
in galactic properties necessary to accurately determine the TF slope.  On the other hand, the magnitude zero point of the relation is best obtained from a sample of distant objects for which a peculiar motion of given amplitude translates into a small magnitude shift.   The SCII sample provides an ideal data set from which to calibrate the TF zero point. 

\subsection{The Accuracy of the Tully-Fisher Zero-Point}
As discussed in Giovanelli et al. (1997b) and Giovanelli et al. (1999), given a number $N$ of clusters the uncertainty on the TF zero point of the resulting template cannot be depressed indefinitely by increasing the average number $\bar n$ of galaxies observed per cluster, and taking advantage of the $\bar n^{-1/2}$ statistical reduction of noise on the mean.  That is because a ``kinematical'' or ``thermal''  component of the uncertainty depends on the number $N$, the distribution in the sky and the peculiar velocity distribution function of the clusters used.  In SCI, for example, the statistical uncertainty deriving from the total number of galaxies observed ($\bar n\times N$) is exceeded by the kinematic uncertainty, which is quantified as follows.  For a sample of $N$ objects with an rms velocity of $\left<V^2\right>^{1/2}$ at a mean redshift of $\left<{\rm c}z\right>$, the expected accuracy of the zero point is limited by systematic concerns to
\begin{equation}
|\Delta m| \approx {2.17 \left<V^2\right>^{1/2} \over \left<{\rm c}z\right> \sqrt{N}} \;\;\; {\rm mag}.
\label {eq:sigma_a}
\end{equation}
This quantity is about 0.04 mag for SCI, while it is only 0.01 mag for SCII due to the larger mean distance and number of clusters of the latter; the SCII sample was selected, in part, to improve upon the accuracy of the kinematical component of the TF zero point.  Since the total number of galaxies involved in the two samples is comparable, the zero point of the SCII template is thus more accurate than that of SCI.  The total uncertainty of the SCII zero point, including the contribution from limitations on its statistical, or internal, accuracy (arising from measurement errors, uncertainties in the corrections applied to observed parameters, the TF scatter, etc) is of order 0.02 magnitudes.

The TF template relation is determined {\it internally} for a cluster sample.
In the case of SCI, it was obtained by assuming that the subset of clusters
farther than 40\h\ Mpc has a globally null monopole (Giovanelli et al. 1997b).  The SCII-based template is obtained by assuming that the
overall set of clusters (wich extends between 50$h^{-1}$ and 200$h^{-1}$ Mpc)
 has a globally null monopole, and adopting the same TF slope as for the SCI sample (Dale et al. 1999c).  This approach does not, however, affect the value of the dipole or higher moments, and thus still allows an effective measure of possible bulk flows.  The SCII template relation is
\begin{equation}
M_I-5\log h = -7.68 (\log W -2.5) - 20.91 \;\; {\rm mag}.
\label{eq:TF}
\end{equation}
The zero points of the SCI and SCII templates were found to agree to within the 
estimated 
uncertainty of 0.02 mag (the SCII zero point is fainter by 0.015 mag).

\subsection{The Tully-Fisher Scatter and its Intrinsic Component}
Any relation involving observed parameters has a limited accuracy described by the amplitude of the relation's scatter.  Claims of the scatter in the TF relation vary from as low as 0.10 mag (Bernstein et al. 1994) to as high as 0.7 mag (Sandage et al. 1994a,b; 1995; Marinoni et al. 1998).  The amplitude of the scatter does depend on wavelength, and studies in the $I$ band typically yield the tightest relations.  The efforts with the largest samples yield 1$\sigma$ dispersion values of $\sim$0.3--0.4 mag (Mathewson, Ford \& Buckhorn 1994; Willick et al. 1995; Giovanelli et al. 1997b; Dale et al. 1999c).  Uncertainties in observational measurements are not the only factors that lead to the overall spread in the data.  The corrections typically applied to the observed fluxes and disk rotational velocities are not exactly known, nor are the variety of methods used to account for inherent sample biases such as cluster incompleteness.  Moreover, there is an {\it intrinsic} component to the TF dispersion since individual galaxies have diverse formation histories.  In fact, Eisenstein \& Loeb (1996) advocate an intrinsic scatter of 0.3 magnitudes, a number greater than most estimates from observational work.  In light of this fact, they make the interesting claim that either spirals formed quite early or that there must be a type of feedback loop that promotes galactic assimilation.
The results of Eisenstein \& Loeb are not corroborated by the results of 
$N$-body simulations (e.g. Baugh et al. 1997, Mo et al. 1998 and Steinmetz
\& Navarro 1999), in which reasonably low values for the TF scatter are
recovered.

\begin{figure}[ht]
\centerline{\psfig{figure=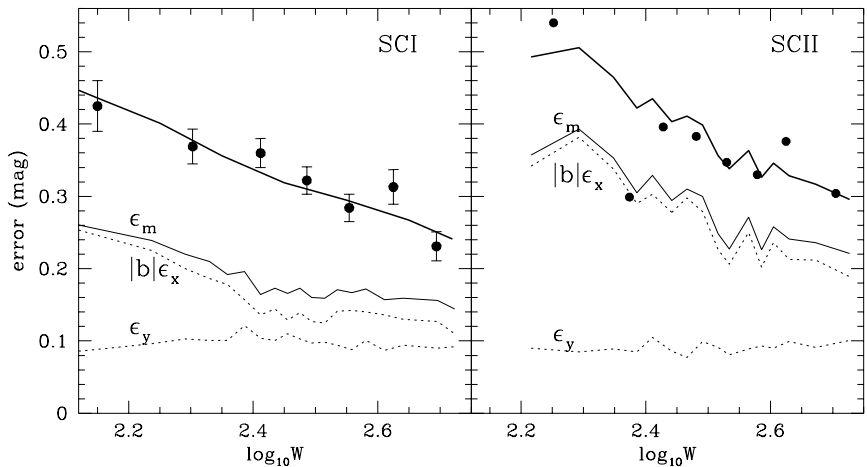,width=4.5in,bbllx=23pt,bblly=195pt,bburx=581pt,bbury=497pt}}
\caption[The Scatter in the Tully-Fisher Relation]
{\ The observed scatter in the TF relation for both the SCI and SCII datasets.  Running averages (in units of magnitudes) of the observed errors are given as dotted lines.  The thin solid line in each panel is the sum in quadrature of the two dotted lines and represents the total 
uncertainty
resulting from measurement errors and that deriving from extinction, geometric
and statistical corrections. 
 The circles plotted indicate the standard deviations of residuals from the template TF relation.  The thick solid line in each panel is the total scatter in the TF relation and has been computed by a sum in quadrature of the observed error distribution and an intrinsic component equivalent to that found in Giovanelli et al. (1997b; see Equation \ref{eq:int} below).}
\label{fig:scatter}
\end{figure}

As already established in Giovanelli et al. (1997b), Dale et al. (1999c), and Willick (1999), Figure \ref{fig:scatter} reinforces the notion of low intrinsic scatter.  
The left panel refers to the SCI data whereas the righthand panel shows the results from SCII.  The two dotted lines in each panel indicate the velocity width and magnitude uncertainties $\epsilon_x$ and $\epsilon_y$, with $\epsilon_x$ multiplied by the TF slope $b$ to put it on a magnitude scale; the thin solid line labeled $\epsilon_m=\sqrt{(b \epsilon_x)^2 + \epsilon_y^2}$ represents the average 
uncertainty resulting from measurement errors and that deriving from extinction, geometric
and statistical corrections. 
The data displayed in Figure \ref{fig:scatter} are generated using equal numbers of galaxies per data point (the difference in the redshift distributions of the SCI and SCII samples result in slightly different observed velocity width distributions).  The circles plotted represent the average standard deviations of the residuals from the fiducial TF relation.  We see that the velocity width errors dominate those from the $I$ band fluxes, which are approximately independent of velocity width.  Furthermore, the logarithmic velocity widths become increasingly uncertain for slower rotators (cf. Giovanelli et al. 1997b; Willick et al. 1997).  We approximate the total observed scatter for the SCI and SCII samples with simple linear relations that depend on the velocity width:
\begin{eqnarray}
\sigma_{\rm tot,SCI} &=& -0.33x + 0.32 \; {\rm mag}.\\
\sigma_{\rm tot,SCII} &=& -0.40x + 0.38 \; {\rm mag}.
\label{eq:scat}
\end{eqnarray}

The total scatter for SCII is in general larger than that for SCI.  This is unsurprising given that SCII velocity widths primarily stem from optical rotation curves instead of 21 cm profiles, a comparatively easier source from which to estimate velocity widths, and since the nearer SCI galaxies generally have better determined disk inclinations.  The gap between the observed scatter and the measured errors is attributed to an intrinsic scatter contribution: the thick top line is a sum in quadrature of our observed measurement errors, $\epsilon_m$, and an intrinsic scatter term (Giovanelli et al. 1997b):
\begin{equation}
\sigma_{\rm int} = -0.28x + 0.26 \; {\rm mag},
\label{eq:int}
\end{equation}
the same in SCI and SCII. This result has been recently confirmed by
Willick and his collaborators, as reported at this meeting.

\section{Peculiar Velocity Distributions}
\begin{figure}
\centerline{\psfig{figure=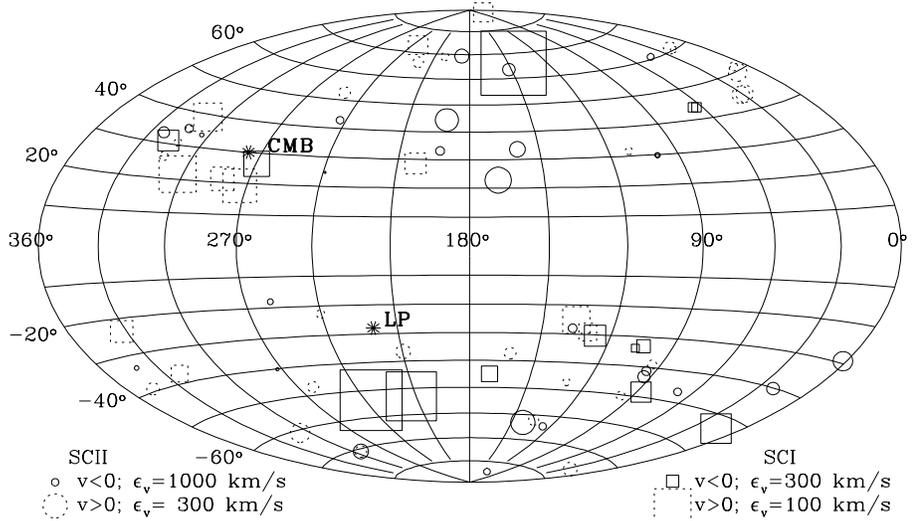,width=12cm,bbllx=33pt,bblly=230pt,bburx=584pt,bbury=560pt}}
\caption[]{The all-sky distribution of the SCI and SCII peculiar velocity samples in Galactic coordinates.  The symbol sizes are inversely proportional to the peculiar velocity uncertainties.  The examples at the bottom give the scales.  Solid/dotted-lined squares refer to positive/negative peculiar velocities in the CMB frame.  Asterisks mark the apices of the motion of the Local Group with respect to the CMB and the Lauer \& Postman (1994) cluster inertial frame.}
\label{fig:aitoff}
\end{figure}

We interpret departures from the canonical template relation as an indication of peculiar motion, with larger departures from the template implying larger amplitude peculiar velocities.  Quantitatively, for 
an
 object at a redshift $z$ with an average 
magnitude departure from the template of 
$\Delta m$, we write the peculiar velocity as 
\begin{equation}
V_{\rm pec} = {\rm c}z \big( 1 - 10^{0.2 \Delta m}).
\label{eq:V}
\end{equation}
Tabulations of the SCI and SCII peculiar motions are given in Giovanelli et al. (1999b) and Dale et a. (1999c), respectively.  Figure \ref{fig:aitoff} shows the distribution of SCI and SCII (CMB frame) peculiar velocities in an Aitoff projection of Galactic Coordinates.  The symbols plotted in the figure reflect both the radial directions of the peculiar velocities and the strengths of the measurements -- in the CMB reference frame, solid/dotted-lined squares represent approaching/receding clusters and the square size is inversely proportional to the accuracy of the measurement.  The largest cluster peculiar velocities, e.g. those for A3266 and A3667, are also the most uncertain, as the clusters are poorly sampled.

\subsection{The RMS One Dimensional Peculiar Velocity Dispersion}

\begin{figure}[!ht]
\centerline{\psfig{figure=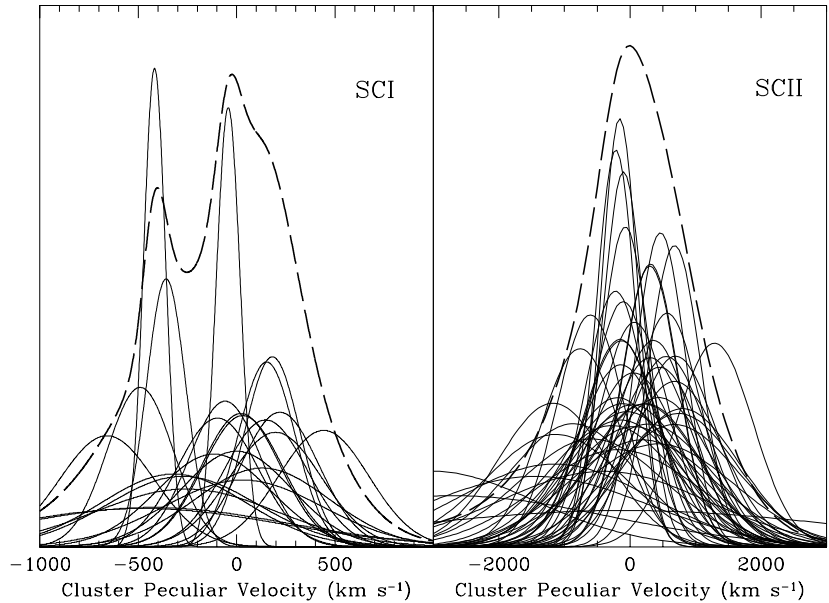,width=4.3in,bbllx=49pt,bblly=195pt,bburx=581pt,bbury=584pt}}
\caption[Peculiar Velocity Distribution]
{\ A display of equal area Gaussians representing the peculiar velocity samples for SCI and SCII.  Each Gaussian is centered at the value of a cluster peculiar velocity, has a dispersion given by the uncertainty in the peculiar velocity measurement, and has an amplitude that is inversely proportional to the uncertainty.  The sum of the Gaussian profiles is given by the thick dashed line.}
\label{fig:gauss_vpecs}
\end{figure}

It is useful to estimate the line-of-sight distribution of peculiar velocities.  The amplitude of that distribution has been known to be a very sensitive discriminator of cosmological models (see, for example, Bahcall \& Oh 1996).  The SCII cluster sample is relatively distant and the cluster membership counts are relatively anemic when compared to SCI.  Consequently, SCII peculiar velocities are much less certain and the overall distribution shown in Figure \ref{fig:gauss_vpecs} is significantly broadened by measurement errors.  The peculiar velocities are represented by equal area Gaussians centered at the peculiar velocity of each cluster with dispersions equal to the estimated peculiar velocity errors.  The thick dashed line superimposed on each plot is the sum of the individual Gaussians (its amplitude has been rescaled for plotting purposes).
The 1$\sigma$ dispersion in the observed distribution of peculiar velocities is found from a Gaussian fit to the dashed line: $\sigma_{1{\rm d,obs,SCI}}=325$ \kms\ and $\sigma_{1{\rm d,obs,SCII}}=796$ \kms.  These values, however, are biased high by measurement errors.  Recovering an estimate of the true values can easily be obtained via Monte Carlo simulations, yielding $\sigma_{\rm 1d,SCI} = 270\pm54$ \kms\ and $\sigma_{\rm 1d,SCII} = 341\pm93$ \kms\ where the error estimates derive from the scatter in the dispersions of the simulated samples.  These values of $\sigma_{\rm 1d}$ are consistent with a relatively low density Universe (Giovanelli et al. 1998b; Bahcall \& Oh 1996; Borgani et al. 1997; Watkins 1999; Bahcall, Gramann \& Cen 1994a; Croft \& Efstathiou 1994).

\section{Dipoles of the Field Spiral Sample (SFI)}
\begin{figure}
\centerline{\psfig{figure=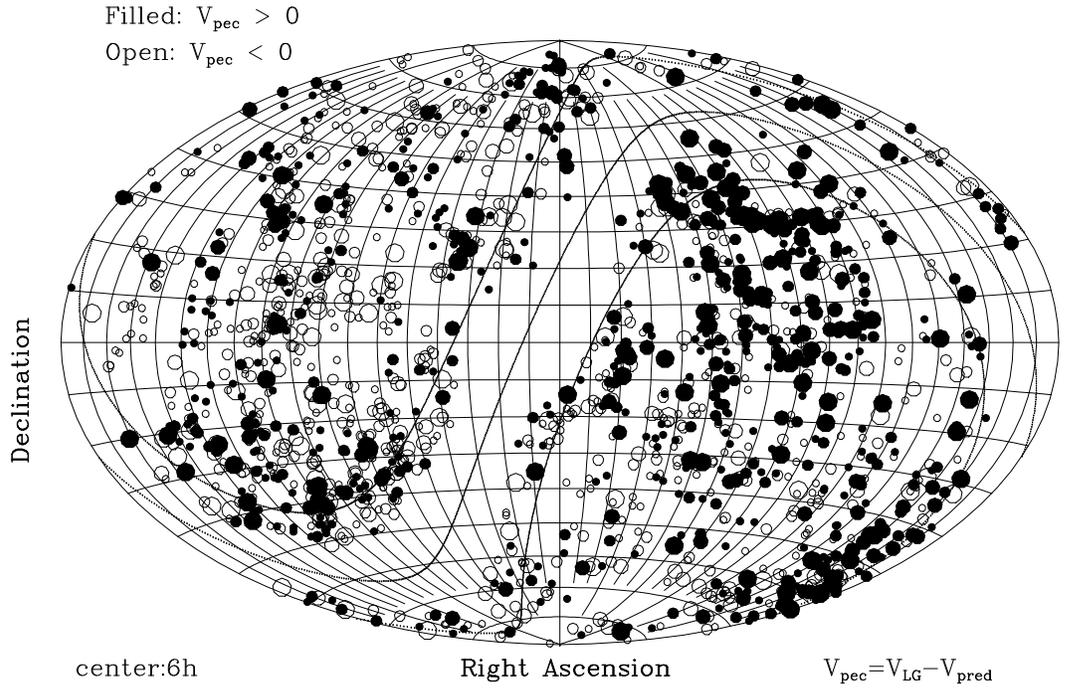,width=14cm,bbllx=48pt,bblly=87pt,bburx=480pt,bbury=748pt,angle=270}}
\caption[]{The distribution of SFI peculiar velocities in the Local Group reference frame.}
\label{fig:all_lg}
\end{figure}

\begin{figure}
\centerline{\psfig{figure=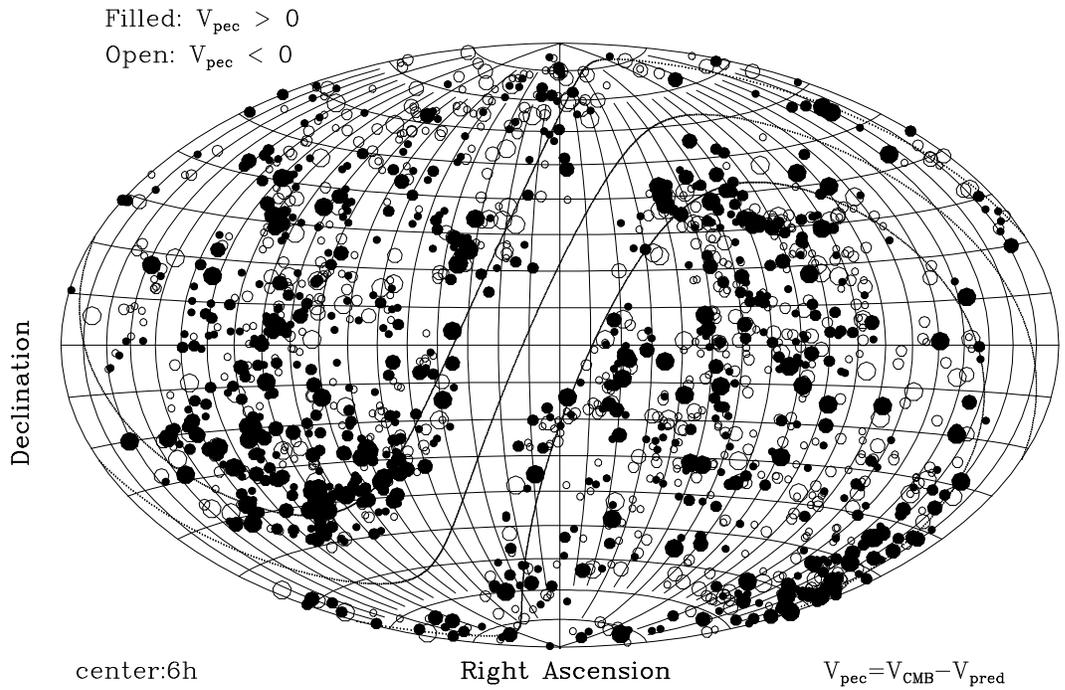,width=14cm,bbllx=48pt,bblly=87pt,bburx=480pt,bbury=748pt,angle=270}}
\caption[]{The distribution of SFI peculiar velocities in the CMB reference frame.}
\label{fig:all_cmb}
\end{figure}

On a grid of equatorial coordinates in an Aitoff projection centered at
$\alpha=6^h$ and $\delta=0^\circ$, Figure \ref{fig:all_lg} displays galaxies from the SFI sample.  Filled and unfilled symbols refer respectively to positive and negative peculiar velocities, measured with respect to the Local Group reference frame.  Lines of Galactic latitude $0^\circ$, $+20^\circ$ and  $-20^\circ$ are also shown, outlining the Zone of Avoidance.  This sample, which extends to 6500 \kms, clearly displays the dipole moment associated with the motion of the Local Group with respect to the CMB, in the form of a prevalence of positive peculiar velocities in the Southern Galactic Hemisphere and of negative ones in the Northern half (the apex of the Local Group motion is near $\alpha=11.2^h$, $\delta=-28^\circ$).  In other words, a large fraction of the volume sampled by the SFI galaxies does not share the Local Group motion resulting in the CMB dipole.  Figure \ref{fig:all_cmb} is a similar display to that in Figure \ref{fig:all_lg}, except that the peculiar velocities are referred to the CMB reference frame.  The dipole signature clear in Figure \ref{fig:all_lg} is gone, indicating that most of the SFI sample has a small bulk flow with respect to the CMB reference frame.  This issue is tackled in a more quantitative way in
what follows.


\begin{figure}
\centerline{\psfig{figure=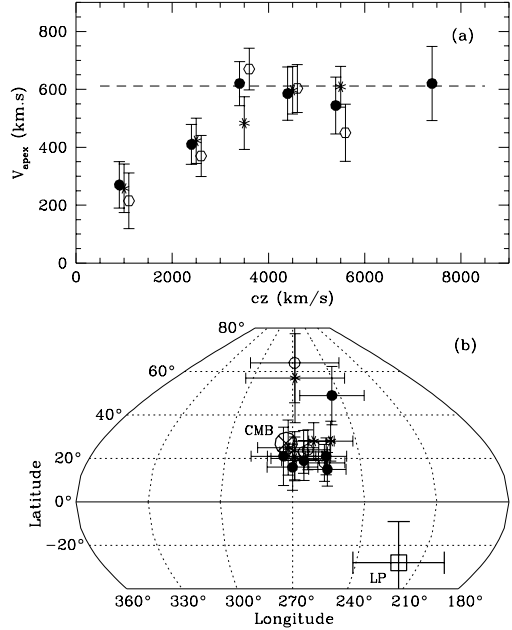,height=8.5cm,bbllx=50pt,bblly=163pt,bburx=460pt,bbury=680pt}}
\caption[]{Amplitudes (a) and apex directions (b) of the peculiar velocity
dipole for the SFI sample. `CMB' and `LP' identify the apices of the CMB and Lauer \& Postman (1994) dipoles.  The dashed line in (a) is the amplitude of the CMB dipole.}
\label{fig:rg2}
\end{figure}

Figure \ref{fig:rg2} displays the dipoles of the reflex motion of the Local Group with respect to field galaxies in shells 2000 \kms\ thick.  The dashed line in panel \ref{fig:rg2}a corresponds to the amplitude of the CMB dipole.  The three sets of symbols identify different ways of computing the peculiar velocities, using different subsets of the data or adopting a direct (stars) or inverse Tully-Fisher relation.  The SFI sample indicates convergence to the CMB dipole, in the motion of the Local Group with respect to galaxies in shells, when a radius of a few thousand \kms\ is reached.  Convergence is achieved both in amplitude and apex direction. 
Note that the reflex motion of the LG is estimated with respect to {\it shells} centered on each redshift, and {\it not} with respect to all galaxies within
that redshift; the convergence rate in the latter case would be slightly slower
than shown in Figure \ref{fig:rg2}.
 The dipole of Lauer \& Postman (1994) is excluded with a high level of confidence.  Alternatively, the bulk flow with respect to the CMB reference frame of a sphere of 6500 \kms\ is 200$\pm$65 \kms, directed toward $(l,b)=(295^\circ,+25^\circ)\pm20^\circ$.  This is in general agreement with the direction of bulk flows reported in other studies (da Costa et al. 1996; Courteau et al. 1993; Dekel 1994; Hudson et al. 1999; Willick 1999), but it is smaller than other determinations, which range between 270 and 700 \kms.  See Giovanelli et al. (1998a) for further details.
Compare these results with the convergence expectations from the PSCz redshift
survey, as presented by W. Saunders at this meeting.

\section{Dipoles of the Cluster Spiral Samples (SCI and SCII)}
\begin{figure}
\centerline{\psfig{figure=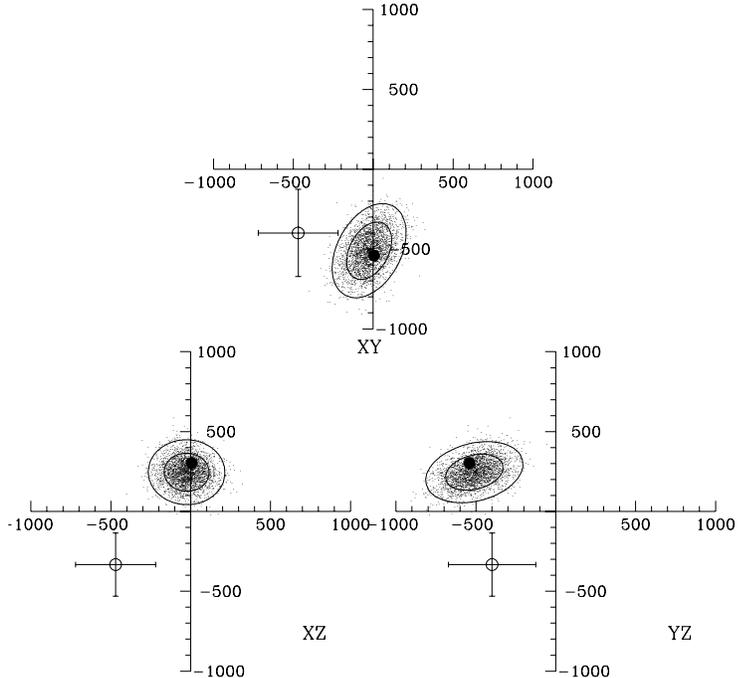,height=9cm,bbllx=15pt,bblly=154pt,bburx=578pt,bbury=675pt}}
\caption[]{Error clouds of the dipole of the reflex motion of the Local Group with respect to the distant SCI+SCII sample. The error clouds (with 1-$\sigma$ and 2-$\sigma$ confidence ellipses) are plotted in supergalactic Cartesian coordinates. The CMB dipole is a filled symbol, which appears enclosed, in all plots, within the 1-$\sigma$ contours; the large cross is the dipole of Lauer \& Postman (1994), which is excluded by the data.}
\label{fig:dd3}
\end{figure}

The dipole moments of the reflex motion of the Local Group with respect to (i) the subset of SCI clusters farther than 3000 \kms\ and (ii) the SCII sample plus the subset of SCI clusters farther than 4500 \kms\ both coincide, within the errors, with that of the CMB.  This coincidence is matched both in amplitude and apex direction.  Figure \ref{fig:dd3} displays the error clouds for the coordinates of the dipole derived from the 
distant SCI+SCII
sample, plotted against supergalactic Cartesian coordinates.  One and two-sigma confidence ellipses are plotted, as well as the locations of the CMB and the Lauer \& Postman (1994) 
reflex
dipoles.  The latter can be excluded by our data with a high level of confidence (at the 3.5$\sigma$ level).  The bulk motion for the 64 clusters comprising the 
distant SCI+SCII
 sample ($\left<{\rm c}z\right> \sim 100h^{-1}$ Mpc) is estimated to be no greater than 200 \kms; corrected for error biasing, the amplitude 
is consistent with
a null bulk flow.  In other words, the inertial frame defined by the 
SCI+SCII
 clusters is 
consistent with
 being at rest with respect to the CMB fiducial rest frame.  More locally, the bulk flow within a sphere of 6000 \kms\ radius, as derived from SCI, is between 140 and 320 \kms, in the CMB reference frame.  See Dale et al. (1999a) and Giovanelli et al. (1998b) for more details.

\section{No Hubble Bubble in the Local Universe}
\begin{figure}
\centerline{\psfig{figure=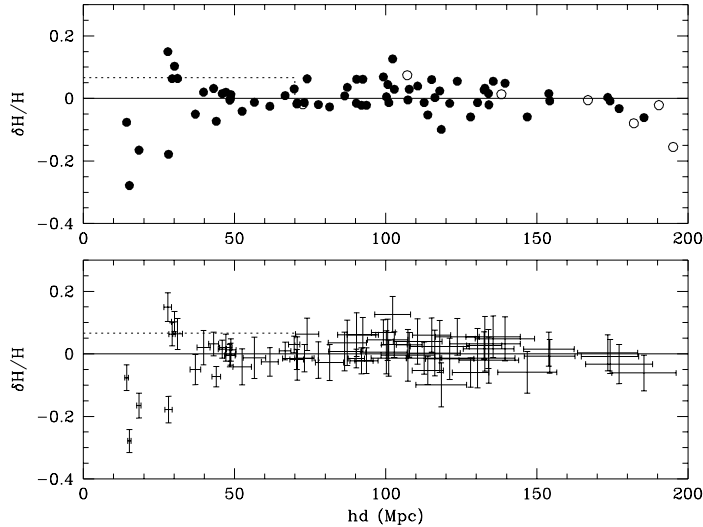,height=7cm,bbllx=50pt,bblly=300pt,bburx=555pt,bbury=680pt}}
\caption[]{No Hubble Bubble in the local Universe. $\delta H/H = 0$ line
identifies null deviation from Hubble flow. The dotted line is the effect
of the Hubble bubble proposed by Zehavi et al. (1998). The lower panel
shows the error bars associated with the filled data points shown above.}
\label{fig:hubbub}
\end{figure}

Recently, it has been suggested by Zehavi et al. (1998) that the volume within 
c$z\simeq 7000$ \kms\ is subject to an acceleration, in the sense that the local Hubble constant is $6.6\pm2.2$\% larger than the fiducial value.  The inference is that we reside at the center of a local underdensity of amplitude 20\%, surrounded by an overdense shell.  The result is based on the distances to 44 SN of type Ia.  In Figure \ref{fig:hubbub}, we test that result by using the combined sample of all 76 SCI+SCII clusters.  The TF distances to each individual cluster have a typical accuracy similar to or better than that quoted for the SN measurements 
(which is 5--8\% due to internal errors alone), 
and the cluster sample is well distributed over the sky (see Figure \ref{fig:aitoff}).  Figure \ref{fig:hubbub} displays $\delta H/H = V_{\rm pec}/{\rm c}z_{\rm tf}$ against $hd = {\rm c}z_{\rm tf}/100$, where $V_{\rm pec}$ is the peculiar velocity in the CMB reference frame and ${\rm c}z_{\rm tf}$ is the Tully-Fisher distance in Mpc, in the same frame.  Unfilled symbols represent clusters with poor distance determinations, based on fewer than 5 galaxies with Tully-Fisher measurements per cluster.  Figure \ref{fig:hubbub} illustrates how, at small distances, the deviations from Hubble flow are dominated by the motions of nearby groups, comparable in amplitude to those of the Local Group (611 \kms).  
At small distances, even modest peculiar velocities constitute a sizable 
fraction of $cz$, thus amplifying and --- if the sampling is sparse ---
even distorting the distribution of values of $\delta H_\circ/H_\circ$.
The deviation from Hubble flow that they imply is of scarce interest.
At distances larger than 35$h^{-1}$ Mpc, the monopole of the cluster peculiar velocity field exhibits no significant change of value.
Our sample is most sensitive to the possible presence of a ``step'' (as would
be produced by the Hubble bubble proposed by Zehavi et al.) between 50$h^{-1}$ and 
130$h^{-1}$ Mpc. The amplitude and significance of a step at 70$h^{-1}$ Mpc is
\begin{equation}
{\delta H_\circ \over H_\circ} = 0.010\pm 0.022
\end{equation}
Our data yield no evidence for a step deviation in the Hubble flow of
amplitude larger than 2--3\%, up to 130$h^{-1}$ Mpc.
The Hubble bubble suggestion of Zehavi et al. (1998) is 
thus
not corroborated by this larger and more accurate data set 
(see Giovanelli et al. 1999 for further details). 

\section {Conclusions}
We have obtained $I$ band Tully-Fisher data for 
3000 spiral galaxies in the field and in 76 clusters, 
evenly spread across the sky 
and distributed between $\sim$10 and 200\h\ Mpc.  Rotational velocity widths derive from either 21 cm or long-slit optical spectroscopy.  The SCI and SCII cluster data are used to construct an accurate Tully-Fisher template relation.  The relation has an average scatter of approximately 0.35 magnitudes, and a zero point with an accuracy of 0.02 mag. 

Peculiar velocities are obtained for 
the SFI galaxies and each of the SCI and SCII clusters,
by reference to the TF template relation. 
The typical distance uncertainty is 
16\% for individual galaxies and 3-9\% for clusters, the latter primarily 
depending on the number of Tully-Fisher measurements within each cluster.  The rms line-of-sight component of the cluster peculiar velocity, debroadened for measurement errors, is $270\pm54$ \kms\ for SCI and $341\pm93$ \kms\ for SCII.
  
Using the data for the field SFI and cluster SCI samples of spiral galaxies, we have shown that the reflex motion of the Local Group converges to the CMB dipole amplitude and direction within 6000 \kms.  Results from the relatively distant SCII cluster sample confirm that convergence is maintained beyond the limits of the SCI and SFI samples.  

Finally, no evidence is found for the local Hubble bubble advocated by Zehavi and coworkers,
or for any radially averaged deviation from Hubble flow with amplitude
larger than 2--3\%, between 35$h^{-1}$ and 130$h^{-1}$ Mpc.
Averaged over distance in excess of a few tens of Mpc, the Hubble flow
appears to be remarkably smooth.

\acknowledgments
The results presented here were obtained by the combined
efforts of M. Haynes, E. Hardy, L. Campusano, M. Scodeggio, J. Salzer, 
G. Wegner, L. da Costa, W. Freudling and the authors.  They are based on 
observations that were carried out at several observatories, including Palomar, 
Arecibo, NOAO, NRAO, Nan\c cay, MPI and MDM. NRAO, NOAO and NAIC are operated 
under management agreements with the National Science Foundation respectively by
AUI, AURA and Cornell University.  The Palomar Observatory is operated by Caltech under a management agreement with Cornell University and JPL.  This research was supported by NSF grants AST94-20505, AST95-28960, and AST96-17069.

\end{document}